\documentclass[onecolumn,showpacs,showkeys,letterpaper]{revtex4}

\usepackage{amsmath}		
\usepackage{amstext}		
\usepackage{amssymb}		
\usepackage{amsfonts}		
\usepackage{graphicx}		
\usepackage{color}		
\usepackage{hyperref}		

\newcommand{\C}[2]{\left(\!\begin{array}{c}#1\\#2\end{array}\!\right)}
\renewcommand{\O}[1]{\mathcal{O}\left(#1\right)}

\newcommand{\diameter}{2}
\newcommand{\node}[1]{\put #1{\circle*{\diameter}}}
\newcommand{\rnode}[1]{\put #1{\circle{\diameter}}}
\newcommand{\cnode}[2]{\put #1{\circle{4}}\put #1{\!\raisebox{-1\unitlength}{#2}}}
\newcommand{\edge}[3]{\put #1{\line #2{#3}}}
\newcommand{\arrow}[4]{\put #1{\line #2{#3}}\put #1{\vector #2{#4}}}
\newcommand{\cadre}[2]{}

\newcommand{\elementaryRoot}{\begin{picture}(2,2)\cadre{2}{2}\rnode{(1,1)}\end{picture}}
\newcommand{\elementaryNode}{\begin{picture}(2,2)\cadre{2}{2}\node{(1,1)}\end{picture}}
\newcommand{\compositeNode}[1]{\begin{picture}(4,3)\cadre{4}{3}\put(2,1){\circle{4}}\put(2,1){\!\raisebox{-1\unitlength}{#1}}\end{picture}}

\newcommand{\CCCa}{\begin{picture}(14,3)\cadre{14}{3}\put(7,1){\oval(14,4)}\node{(2,1)}\edge{(2,1)}{(1,0)}{5}\node{(7,1)}\edge{(7,1)}{(1,0)}{5}\node{(12,1)}\end{picture}}

\newcommand{\Ga}{\begin{picture}(2,2)\cadre{2}{2}\node{(1,1)}\end{picture}}
\newcommand{\GGa}{\begin{picture}(7,2)\cadre{7}{2}\node{(1,1)}\edge{(1,1)}{(1,0)}{5}\node{(6,1)}\end{picture}}
\newcommand{\GGGa}{\begin{picture}(12,2)\cadre{12}{2}\node{(1,1)}\edge{(1,1)}{(1,0)}{5}\node{(6,1)}\edge{(6,1)}{(1,0)}{5}\node{(11,1)}\end{picture}}
\newcommand{\GGGb}{\begin{picture}(4,3)\cadre{4}{3}\cnode{(2,1)}{3}\end{picture}}
\newcommand{\GGGGa}{\begin{picture}(17,2)\cadre{17}{2}\node{(1,1)}\edge{(1,1)}{(1,0)}{5}\node{(6,1)}\edge{(6,1)}{(1,0)}{5}\node{(11,1)}\edge{(11,1)}{(1,0)}{5}\node{(16,1)}\end{picture}}
\newcommand{\GGGGb}{\begin{picture}(10,5)\cadre{10}{5}\node{(1,1)}\edge{(1,1)}{(1,0)}{5}\node{(6,1)}\edge{(6,1)}{(1,1)}{3.7}\node{(9,4)}\edge{(6,1)}{(1,-1)}{3.7}\node{(9,-2)}\end{picture}}
\newcommand{\GGGGc}{\begin{picture}(9,3)\cadre{9}{3}\cnode{(2,1)}{3}\edge{(3.9,1)}{(1,0)}{4}\node{(8,1)}\end{picture}}
\newcommand{\GGGGGa}{\begin{picture}(22,2)\cadre{22}{2}\node{(1,1)}\edge{(1,1)}{(1,0)}{5}\node{(6,1)}\edge{(6,1)}{(1,0)}{5}\node{(11,1)}\edge{(11,1)}{(1,0)}{5}\node{(16,1)}\edge{(16,1)}{(1,0)}{5}\node{(21,1)}\end{picture}}
\newcommand{\GGGGGb}{\begin{picture}(15,5)\cadre{15}{5}\node{(1,1)}\edge{(1,1)}{(1,0)}{5}\node{(6,1)}\edge{(6,1)}{(1,0)}{5}\node{(11,1)}\edge{(11,1)}{(1,1)}{3.7}\node{(14,4)}\edge{(11,1)}{(1,-1)}{3.7}\node{(14,-2)}\end{picture}}
\newcommand{\GGGGGc}{\begin{picture}(12,7)\cadre{12}{7}\node{(1,1)}\edge{(1,1)}{(1,0)}{5}\node{(6,1)}\edge{(6,1)}{(1,0)}{5}\node{(11,1)}\edge{(6,1)}{(0,1)}{5}\node{(6,6)}\edge{(6,1)}{(0,-1)}{5}\node{(6,-4)}\end{picture}}
\newcommand{\GGGGGd}{\begin{picture}(14,3)\cadre{14}{3}\cnode{(2,1)}{3}\edge{(3.9,1)}{(1,0)}{4}\node{(8,1)}\edge{(8,1)}{(1,0)}{5}\node{(13,1)}\end{picture}}
\newcommand{\GGGGGe}{\begin{picture}(14,3)\cadre{14}{3}\node{(1,1)}\edge{(1,1)}{(1,0)}{4}\cnode{(7,1)}{3}\edge{(8.9,1)}{(1,0)}{4}\node{(13,1)}\end{picture}}
\newcommand{\GGGGGf}{\begin{picture}(4,3)\cadre{4}{3}\cnode{(2,1)}{5}\end{picture}}
\newcommand{\GGGGGGc}{\begin{picture}(13,5)\cadre{13}{5}\node{(1,-2)}\edge{(1,-2)}{(1,1)}{3.7}\node{(1,4)}\edge{(1,4)}{(1,-1)}{3.7}\node{(4,1)}\edge{(4,1)}{(1,0)}{5}\node{(9,1)}\edge{(9,1)}{(1,1)}{3.7}\node{(12,4)}\edge{(9,1)}{(1,-1)}{3.7}\node{(12,-2)}\end{picture}}
\newcommand{\GGGGGGi}{\begin{picture}(11,5)\cadre{11}{5}\node{(1,-2)}\edge{(1,-2)}{(1,1)}{3.7}\node{(1,4)}\edge{(1,4)}{(1,-1)}{3.7}\node{(4,1)}\edge{(4,1)}{(1,0)}{3}\cnode{(9,1)}{3}\end{picture}}

\newcommand{\tildeGa}{\begin{picture}(2,2)\cadre{2}{2}\rnode{(1,1)}\end{picture}}
\newcommand{\tildeGGa}{\begin{picture}(4,5)\cadre{4}{5}\rnode{(1,4)}\arrow{(1,3)}{(0,-1)}{5}{3}\put(1.5,0){$i_{1}$}\node{(1,-2)}\end{picture}}
\newcommand{\tildeGGaa}{\begin{picture}(4,5)\cadre{4}{5}\rnode{(1,4)}\arrow{(1,-2)}{(0,1)}{5}{4}\put(1.5,0){$i_{1}$}\node{(1,-2)}\end{picture}}
\newcommand{\tildeGGGa}{\begin{picture}(10,5)\cadre{10}{5}\rnode{(5,4)}\arrow{(4.55279,3.10557)}{(-1,-2)}{3}{1.7}\arrow{(5.44721,3.10557)}{(1,-2)}{3}{1.7}\put(0,0){$i_{1}$}\put(7.5,0){$i_{2}$}\node{(2,-2)}\node{(8,-2)}\end{picture}}
\newcommand{\tildeGGGb}{\begin{picture}(11,6)\cadre{11}{6}\put(5.5,0.8){\circle{10.6}}\rnode{(5.5,4.5)}\arrow{(5.05279,3.60557)}{(-1,-2)}{3}{1.7}\arrow{(5.94721,3.60557)}{(1,-2)}{3}{1.7}\put(0.5,0.5){$i_{1}$}\put(8,0.5){$i_{2}$}\node{(2.5,-1.5)}\node{(8.5,-1.5)}\end{picture}}
\newcommand{\tildeGGGGa}{\begin{picture}(11,8)\cadre{11}{8}\rnode{(5,7)}\arrow{(4.55279,6.10557)}{(-1,-2)}{3}{1.7}\arrow{(5.44721,6.10557)}{(1,-2)}{3}{1.7}\put(0,3){$i_{1}$}\put(7.5,3){$i_{2}$}\node{(2,1)}\node{(8,1)}\arrow{(8,1)}{(0,-1)}{6}{4}\put(8.5,-3){$i_{3}$}\node{(8,-5)}\end{picture}}
\newcommand{\tildeGGGGb}{\begin{picture}(10,8)\cadre{10}{8}\rnode{(5,7)}\arrow{(5,6)}{(0,-1)}{5}{3}\put(5.5,3){$i_{1}$}\node{(5,1)}\arrow{(5,1)}{(-1,-2)}{3}{2}\arrow{(5,1)}{(1,-2)}{3}{2}\put(0,-3){$i_{2}$}\put(7.5,-3){$i_{3}$}\node{(2,-5)}\node{(8,-5)}\end{picture}}
\newcommand{\tildeGGGGc}{\begin{picture}(12.5,9.5)\cadre{12.5}{9.5}\qbezier(4,4)(-1,14)(10,6)\qbezier(10,6)(13.0152,3.80711)(12,0)\qbezier(12,0)(8,-15)(5,0)\qbezier(5,0)(4.33333,3.33333)(4,4)\rnode{(5,7)}\arrow{(4.55279,6.10557)}{(-1,-2)}{3}{1.7}\arrow{(5.44721,6.10557)}{(1,-2)}{3}{1.7}\put(0,3){$i_{1}$}\put(7.5,3){$i_{2}$}\node{(2,1)}\node{(8,1)}\arrow{(8,1)}{(0,-1)}{6}{4}\put(8.5,-3){$i_{3}$}\node{(8,-5)}\end{picture}}
\newcommand{\tildeGGGGd}{\begin{picture}(11,8)\cadre{11}{8}\put(5.5,-2.7){\circle{10.6}}\rnode{(5.5,7)}\arrow{(5.5,6)}{(0,-1)}{6}{3}\put(6.2,3.5){$i_{1}$}\node{(5.5,1)}\arrow{(5.5,1)}{(-1,-2)}{3}{2}\arrow{(5.5,1)}{(1,-2)}{3}{2}\put(0.5,-3){$i_{2}$}\put(8,-3){$i_{3}$}\node{(2.5,-5)}\node{(8.5,-5)}\end{picture}}
\newcommand{\tildeGGGGGGGGGy}{\begin{picture}(15,14)\cadre{15}{14}\put(5.5,8.8){\circle{10.6}}\rnode{(5.5,12.5)}\arrow{(5.05279,11.60557)}{(-1,-2)}{3}{1.7}\arrow{(5.94721,11.60557)}{(1,-2)}{3}{1.7}\put(0.5,8.5){$i_{1}$}\put(8,8.5){$i_{2}$}\node{(2.5,6.5)}\node{(8.5,6.5)}\arrow{(2.5,6.5)}{(0,-1)}{6}{4}\arrow{(8.5,6.5)}{(0,-1)}{6}{4}\put(-0.4,2.5){$i_{3}$}\put(9.3,2.6){$i_{4}$}\node{(2.5,0.5)}\put(8.5,-3.2){\circle{10.6}}\node{(8.5,0.5)}\arrow{(5.5,-5.5)}{(+1,+2)}{3}{2}\arrow{(8.5,0.5)}{(1,-2)}{3}{2}\put(3.5,-3.3){$i_{5}$}\put(10.8,-3.3){$i_{6}$}\node{(5.5,-5.5)}\node{(11.5,-5.5)}\arrow{(5.5,-5.5)}{(0,-1)}{6}{4}\arrow{(11.5,-11.5)}{(0,+1)}{6}{4}\put(2.4,-9.5){$i_{7}$}\put(12.1,-9.6){$i_{8}$}\node{(5.5,-11.5)}\node{(11.5,-11.5)}\end{picture}}
\newcommand{\tildeGGGGGGGGGz}{\begin{picture}(17,10)\cadre{17}{10}\qbezier(1.5,7)(8.5,12)(15.5,7)\qbezier(15.5,7)(19,4.5)(13.5,1.2)\qbezier(13.5,1.2)(8.5,-1.8)(3.5,1.2)\qbezier(3.5,1.2)(-2,4.5)(1.5,7)\rnode{(8.5,1)}\arrow{(9.39443,1.44721)}{(2,1)}{6}{4}\arrow{(8.94721,1.89443)}{(1,2)}{3}{2}\arrow{(8.05279,1.89443)}{(-1,2)}{3}{2}\arrow{(7.60557,1.44721)}{(-2,1)}{6}{4}\arrow{(7.60557,0.55279)}{(-2,-1)}{6}{4}\arrow{(8.05279,0.10557)}{(-1,-2)}{3}{2}\arrow{(8.94721,0.10557)}{(1,-2)}{3}{2}\arrow{(9.39443,0.55279)}{(2,-1)}{6}{4}\put(4,3.6){\tiny$i_{1}$}\put(6.3,5.5){\tiny$i_{2}$}\put(8.4,5.9){\tiny$i_{5}$}\put(11.1,4.1){\tiny$i_{6}$}\put(3.6,-2.7){\tiny$i_{3}$}\put(6.3,-4.5){\tiny$i_{4}$}\put(8.6,-4.5){\tiny$i_{7}$}\put(10.8,-2.2){\tiny$i_{8}$}\node{(14.5,4)}\node{(11.5,7)}\node{(5.5,7)}\node{(2.5,4)}\node{(2.5,-2)}\node{(5.5,-5)}\node{(11.5,-5)}\node{(14.5,-2)}\end{picture}}

\begin{document}

\title{A combinatorial solution for the current fluctuations in the exclusion process}
\author{Sylvain Prolhac}
\email[]{sylvain.prolhac@cea.fr}
\affiliation{Institut de Physique Th\'eorique,\\
CEA, IPhT, F-91191 Gif-sur-Yvette, France\\
CNRS, URA 2306, F-91191 Gif-sur-Yvette, France}
\date{July 31, 2009}

\begin{abstract}
We conjecture an exact expression for the large deviation function of the stationary state current in the partially asymmetric exclusion process with periodic boundary conditions. This expression is checked for small systems using functional Bethe Ansatz. It generalizes a previous result by Derrida and Lebowitz for the totally asymmetric exclusion process, and gives the known values for the three first cumulants of the current in the partially asymmetric model. Our result is written in terms of tree structures and provides a new example of a link between integrable models and combinatorics.

\end{abstract}

\maketitle

\begin{section}{Introduction}
The asymmetric simple exclusion process (ASEP) is one of the simplest interacting particles systems featuring an out of equilibrium stationary state. It has been studied much in the past \cite{S91.1,HHZ95.1,SZ98.1}, in particular because it belongs to the class of exactly solvable models. A quantity of interest is the macroscopic stationary state current and its fluctuations, since the presence of this current is the signature that the system is out of equilibrium. Various boundary conditions have been used in the study of the one-dimensional ASEP: open boundaries connecting the system to reservoirs of particles \cite{D07.1,DLS02.1,dGE05.1}, infinite line $\mathbb{Z}$ \cite{S06.1,S07.1,RS06.1}, and periodic boundary conditions \cite{GM06.1,DE99.1,P03.1}, which is the case studied here.\\\indent
Using the Bethe Ansatz, all the cumulants of the current were calculated \cite{DL98.1} in the special case of the totally asymmetric simple exclusion process (TASEP), for which the particles hop in only one direction. For an arbitrary asymmetry between the hopping rates, finite size expressions for the three first cumulants were derived \cite{PM08.1,P08.1}, and all the cumulants were obtained \cite{LK99.1} in the large system size limit with non vanishing asymmetry.\\\indent
In the present work, we conjecture an expression (\ref{E(gamma) forests}) generalizing these results: it provides an exact expression for all the cumulants of the stationary state current in the ASEP with partial asymmetry for finite systems. This expression gives the TASEP result \cite{DL98.1} and the limit obtained in \cite{LK99.1}. It also allows the study of a vanishing asymmetry probing the transition between the equilibrium system with symmetric rates and the totally asymmetric system for which detailed balance is maximally broken. The exact formulas obtained previously for the three first cumulants in the partially asymmetric model are also recovered. Our conjecture is checked for small systems using functional Bethe Ansatz.\\\indent
We consider the asymmetric simple exclusion process on a ring of size $L$ with $n$ particles hopping locally both one site forward (with rate $p$) and backward (with rate $q=xp$). By the exclusion rule, the particles are only allowed to hop if the destination site is empty. The integrated current $Y_{t}$ is defined as the total distance covered by all the particles between time $0$ and time $t$. When $t$ becomes large, the fluctuations of $Y_{t}$ are given by \cite{DL98.1,PM08.1}
\begin{equation}
\langle e^{\gamma Y_{t}}\rangle\sim e^{E(\gamma)t}\;.
\end{equation}
The formal series $E(\gamma)$ is the exponential generating function of the cumulants of the current:
\begin{equation}
E(\gamma)=J\gamma+\frac{D}{2!}\gamma^{2}+\frac{E_{3}}{3!}\gamma^{3}+\ldots\;,
\end{equation}
where $J$ is the mean value of the current, $D$ the diffusion constant and $E_{3}$ the third cumulant of the current. The generating function $E(\gamma)$, which is related to the large deviation function of the current by a Legendre transform, is also the eigenvalue with largest real part of a deformation $M(\gamma)$ of the Markov matrix of the system \cite{DL98.1,PM08.1}. The matrix $M(\gamma)$ is similar to the hamiltonian of a XXZ spin chain with twisted boundary conditions \cite{GM06.1}, and can thus be diagonalized using the Bethe Ansatz. The generating function of the cumulants of the current is given by
\begin{equation}
\label{E[Q]}
\frac{E(\gamma)}{p}=(1-x)\frac{d}{dt}\log\left(\frac{Q(t)}{x^{n}Q(t/x)}\right)_{|t=1}\;,
\end{equation}
where the polynomial $Q$ of degree $n$, along with a polynomial $R$ of degree $L$, is a solution of the functional Bethe equation \cite{B82.1}
\begin{equation}
\label{ebQR}
Q(t)R(t)=e^{L\gamma}(1-t)^{L}Q(xt)+(1-xt)^{L}x^{n}Q(t/x)\;.
\end{equation}
This equation has several solutions, corresponding to different eigenstates of the matrix $M(\gamma)$. The solution of equation (\ref{ebQR}) corresponding to the largest eigenvalue of $M(\gamma)$ verifies \cite{PM08.1}
\begin{equation}
\label{Q(gamma=0)}
Q(t)=t^{n}+\O{\gamma}\;,
\end{equation}
and
\begin{equation}
\label{Q(1)}
e^{n\gamma}Q(1)=x^{n}Q(1/x)\;.
\end{equation}
Using (\ref{Q(gamma=0)}) and (\ref{Q(1)}), the functional Bethe equation (\ref{ebQR}) can be solved perturbatively near $\gamma=0$ \cite{P08.1}.
\end{section}

\begin{section}{Tree structures}
We now introduce a few combinatorial structures in terms of which the generating function of the cumulants of the current $E(\gamma)$ will be expressed. We call ``composite node'' a (finite) set containing a strictly positive odd number of ``elementary nodes''. The size $|c|$ of a composite node $c$ is defined to be the number of elementary nodes it contains. In the following, two different sets of composite nodes will be considered: the set $\mathcal{C}$ of composite nodes without internal structure and the set $\widetilde{\mathcal{C}}$ of composite nodes with an internal unrooted tree structure \cite{FS09.1} (or acyclic graph structure) linking all the elementary nodes it contains. The elementary nodes and the composite nodes of size $1$ will both be represented by a dot (\elementaryNode), the composite nodes elements of $\mathcal{C}\backslash\{\elementaryNode\}$ by their size surrounded by a circle (\textit{e.g.} \compositeNode{3}, \compositeNode{5}, \ldots), and the composite nodes elements of $\widetilde{\mathcal{C}}\,\backslash\{\elementaryNode\}$ by the tree structure on the elementary nodes they contain surrounded by a closed line (\textit{e.g.} \CCCa).\\\indent
We can now build trees whose nodes will be composite nodes. We call $\mathcal{G}$ the set of unrooted trees with nodes elements of $\mathcal{C}$ and $\widetilde{\mathcal{G}}$ the set of rooted trees with nodes elements of $\widetilde{\mathcal{C}}$ and oriented edges labeled by $i_{1}$, $i_{2}$, \ldots One arbitrary elementary node of $g\in\widetilde{\mathcal{G}}$, chosen as the root of $g$, will be represented by a small circle (\elementaryRoot). The edges between composite nodes are identified to edges between elementary nodes belonging to different composite nodes. These edges will be called ``outer edges'', by opposition to the ``inner edges'' linking elementary nodes contained in the same composite node. For $g\in\mathcal{G}$ or $g\in\widetilde{\mathcal{G}}$, we define the size $|g|$ of $g$ as the sum of the sizes of the composite nodes of $g$. We call $\mathcal{G}_{r}$ (respectively $\widetilde{\mathcal{G}}_{r}$) the subset of $\mathcal{G}$ (resp. $\widetilde{\mathcal{G}}$) consisting of trees of size $r$. The first sets $\mathcal{G}_{r}$ and $\widetilde{\mathcal{G}}_{r}$ are drawn in fig.\ref{fig G} and fig.\ref{fig Gtilde}.\\\indent
\begin{figure}[ht]
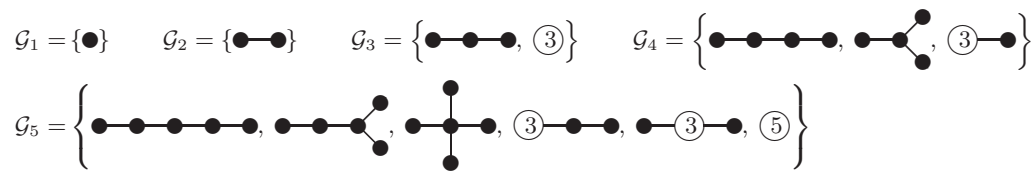

\begin{align*}
\mathcal{G}_{1}=&\left\{\Ga\right\}\qquad\mathcal{G}_{2}=\left\{\GGa\right\}\qquad\mathcal{G}_{3}=\left\{\GGGa,\;\GGGb\right\}\qquad\mathcal{G}_{4}=\left\{\GGGGa,\;\GGGGb,\;\GGGGc\right\}\\
\mathcal{G}_{5}=&\left\{\GGGGGa,\;\GGGGGb,\;\GGGGGc,\;\GGGGGd,\;\GGGGGe,\;\GGGGGf\right\}
\end{align*}
\caption{The five first $\mathcal{G}_{r}$ sets.}
\label{fig G}
\end{figure}
\begin{figure}[ht]
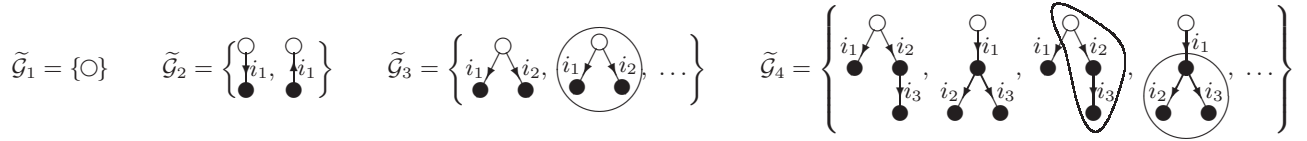

\begin{equation*}
\widetilde{\mathcal{G}}_{1}=\left\{\tildeGa\right\}\qquad\widetilde{\mathcal{G}}_{2}=\left\{\tildeGGa,\;\tildeGGaa\right\}\qquad\widetilde{\mathcal{G}}_{3}=\left\{\tildeGGGa,\;\tildeGGGb,\;\ldots\right\}\qquad\widetilde{\mathcal{G}}_{4}=\left\{\tildeGGGGa,\;\tildeGGGGb,\;\tildeGGGGc,\;\tildeGGGGd,\;\ldots\right\}
\end{equation*}
\caption{The four first $\widetilde{\mathcal{G}}_{r}$ sets. The $\ldots$ represent trees obtained from the trees drawn by changing the root, the direction of the arrows on the edges and doing any permutation of the labels of the edges.}
\label{fig Gtilde}
\end{figure}
For $g\in\mathcal{G}$, we call $c(g)$ the set of the composite nodes of $g$. For a composite node $c\in c(g)$, we define $v_{c}$ as the number of composite nodes in $g$ which are neighbors of $c$. For $g\in\widetilde{\mathcal{G}}$, $e(g)$ will be the set of the elementary nodes of $g$ and $o(g)$ the set of the outer edges of $g$. For an outer edge $o\in o(g)$, we call $\ell(o)$ the label of $o$ and $m(o)$ the sum of the labels of $g$ which label both the inner and outer edges of the subtree of $g$ beginning at $o$, that is, the sum of all the labels that can be attained from $o$ by moving on the edges of $g$ away from the root of $g$. Finally, for an elementary node $e$ of $g\in\widetilde{\mathcal{G}}$, we will also use the notation $\ell(e)$ for the sum of the labels of the edges pointing to $e$ minus the sum of the labels of the edges leaving $e$.\\\indent
For $g\in\widetilde{\mathcal{G}}$, we define a tree $g^{*}\in\widetilde{\mathcal{G}}$ as follows: we attach all the elementary nodes of $g$ to the root such that all the edges point away from the root and keep their labels. Then, we set the composite nodes such that all the composite nodes are of size $1$ except the one containing the root, and such that the labels of the outer edges of $g^{*}$ are the same as the labels of the outer edges of $g$. For example, we have
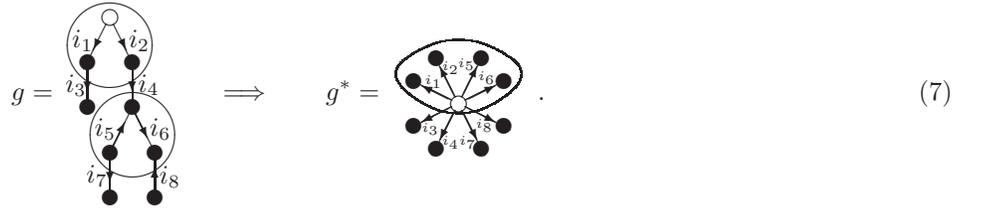
\begin{equation}
\label{example g g*}
g=\begin{array}{c}\tildeGGGGGGGGGy\\\\\\\end{array}\quad\Longrightarrow\qquad g^{*}=\begin{array}{c}\tildeGGGGGGGGGz\\\\\end{array}\;.
\end{equation}
For these trees, the values $\ell(e)$ for the elements $e$ of $e(g)$ and $e(g^{*})$ are respectively
\begin{align}
g:\;\;&i_{3},\,i_{1}-i_{3},\,-i_{1}-i_{2},\,i_{2}-i_{4},\,i_{4}+i_{5}-i_{6},\,-i_{5}-i_{7},\,i_{7},\,i_{6}+i_{8},\,-i_{8}\;.\\
g^{*}:\;&i_{1},\,i_{2},\,i_{3},\,i_{4},\,i_{5},\,i_{6},\,i_{7},\,i_{8},\,-i_{1}-i_{2}-i_{3}-i_{4}-i_{5}-i_{6}-i_{7}-i_{8}\;,
\end{align}
while the values $m(o)$ for the elements $o$ of $o(g)$ and $o(g^{*})$ are given by
\begin{align}
g:\;\;&i_{3},\,i_{7},\,i_{8},\,i_{4}+i_{5}+i_{6}+i_{7}+i_{8}\;.\\
g^{*}:\;&i_{3},\,i_{4},\,i_{7},\,i_{8}\;.
\end{align}
We will now define some functions acting on $\mathcal{G}$ and $\widetilde{\mathcal{G}}$. For $g\in\widetilde{\mathcal{G}}$ and two arbitrary functions $\varphi$ and $\eta$, we define
\begin{equation}
\label{U(g)}
U_{\varphi,\eta}(g)=\left(\sum_{e\in e(g)}\varphi(\ell(e))\right)\left(\prod_{e\in e(g)}\eta(\ell(e))\right)\;.
\end{equation}
It can be shown that $U_{\varphi,\eta}(g)$ does not depend on the position of the root or on the composite nodes of $g$, but only on the tree structure of the elementary nodes of $g$. For an arbitrary even function $\xi$, we also define
\begin{equation}
\label{V(g)}
V_{\xi}(g)=\prod_{o\in o(g)}\xi(m(o))\;.
\end{equation}
Unlike $U_{\varphi,\eta}(g)$, $V_{\xi}(g)$ does depend on the position of the root and of the composite nodes in $g$, but not on the direction of the edges and on the internal tree structure of the composite nodes. We choose an arbitrary application $\theta$ from $\mathcal{G}$ to $\widetilde{\mathcal{G}}$ preserving the tree structure on the composite nodes, which roots the trees, adds an internal tree structure on the composite nodes, labels the edges, and adds arrows to them. Then, we define for $g\in\mathcal{G}_{r}$
\begin{equation}
\label{W(g) tree 1}
W_{\varphi}^{\eta,\xi}(g)=\sum_{i_{1}\in\mathbb{Z}}\ldots\!\sum_{i_{r-1}\in\mathbb{Z}}U_{\varphi,\eta}(\theta(g))V_{\xi}(\theta(g)^{*})\;.
\end{equation}
For $r=1$, we take $W_{\varphi}^{\eta,\xi}(\Ga)=\varphi(0)\eta(0)$. It can be proved that $W_{\varphi}^{\eta,\xi}$ does not depend on the choice of the function $\theta$. Performing some changes of variables on the summation indices $i_{1}$, \ldots, $i_{r-1}$ in equation (\ref{W(g) tree 1}), an equivalent expression for $W_{\varphi}^{\eta,\xi}(g)$ can be written:
\begin{equation}
\label{W(g) tree 2}
W_{\varphi}^{\eta,\xi}(g)=\sum_{i_{1}\in\mathbb{Z}}\ldots\!\sum_{i_{r-1}\in\mathbb{Z}}U_{\varphi,\eta}(\theta(g)^{*})V_{\xi}(\theta(g))\;.
\end{equation}
For a tree $g\in\mathcal{G}$, we define a symmetry factor $S_{t}(g)$ associated to $g$ by
\begin{equation}
S_{t}(g)=P_{t}(g)\prod_{c\in c(g)}(-1)^{\frac{|c|-1}{2}}\frac{|c|^{3}|c|!}{|c|^{v_{c}}(|c|!!)^{2}}\;,
\end{equation}
where $P_{t}(g)$ is equal to the number of permutations of the composite nodes of $g$ which leave it invariant. Table \ref{table symmetry factors} gives a few examples of symmetry factors. These symmetry factors can be used to define a suitable generating function for the trees of $\mathcal{G}$. From a direct calculation up to $r=16$, we observed that the following generating function can be expressed in a simple closed form:
\begin{equation}
\label{GF trees}
Z_{r}(z)=\sum_{g\in\mathcal{G}_{r}}\frac{z^{|c(g)|}}{S_{t}(g)}=\frac{z}{r\times r!}\prod_{j=1}^{r-1}\left[r(z+1)-2j\right]\;,
\end{equation}
where $|c(g)|$ is the number of composite nodes in $g$.
\begin{table}[ht]
\begin{tabular}{cccccccc}
$g\in\mathcal{G}$ & \;\Ga & \;\GGa & \;\GGGa & \;\GGGb & \;\GGGGGf & \;\GGGGGGc & \;\GGGGGGi\\\\
$S_{t}(g)$ & 1 & 2 & 2 & -18 & 200/3 & 8 & -12\\\\
$P_{t}(g)$ & 1 & 2 & 2 & 1 & 1 & 8 & 2
\end{tabular}
\caption{Examples of symmetry factors of trees $g\in\mathcal{G}$}
\label{table symmetry factors}
\end{table}
\end{section}

\begin{section}{Parametric expression for the current fluctuations}
We now state our conjecture about the current fluctuations for the ASEP in terms of the trees and forests that we introduced before. We define the function $\varphi_{l}$ by
\begin{equation}
\varphi_{l}(z)=\frac{(n+z)}{(L-l)}\C{L-n-z}{l}\left/\C{L}{l}\right.\;,
\end{equation}
the function $\eta$ by
\begin{equation}
\eta(z)=\C{L}{n+z}\left/\C{L}{n}\right.\;,
\end{equation}
and the (even) function $\xi_{x}$ by
\begin{equation}
\xi_{x}(z)=\left\{\begin{array}{cl}1 & \text{if $z=0$}\\\frac{1+x^{|z|}}{1-x^{|z|}} & \text{if $z\neq 0$}\end{array}\right.\;.
\end{equation}
The value $\xi_{x}(0)$ was set to $1$, but we emphasize that an arbitrary value could have been taken by modifying the following accordingly. With these definitions, the polynomial $Q$ solution of (\ref{ebQR}--\ref{Q(1)}) is then given by
\begin{equation}
\label{A(t) trees}
\ln\left[\frac{Q(t)}{x^{n}Q(t/x)}\right]=\sum_{k=1}^{\infty}\sum_{l=0}^{\infty}\frac{B^{k}(1-t)^{l}}{2^{k-1}}\sum_{g\in\mathcal{G}_{k}}\frac{W_{\varphi_{l}}^{\eta,\xi_{x}}(g)}{S_{t}(g)}\;,
\end{equation}
with
\begin{equation}
\label{B[Q(0)]}
B=(-1)^{n-1}\C{L}{n}\left(e^{L\gamma}-x^{n}\right)Q(0)\;.
\end{equation}
Using the method developed in \cite{P08.1} to solve the functional Bethe equation (\ref{ebQR}) perturbatively up to order $5$ in $\gamma$ and $1-t$, we checked equations (\ref{A(t) trees}) and (\ref{B[Q(0)]}) at this order for all the systems up to size $12$. Making the analytic continuation for complex $L$ in equation (\ref{A(t) trees}), the apparent divergences from $\varphi_{l}$ when $L$ takes an integer value such that $L\leq l$ vanish since $\varphi_{l}(z)$ is always multiplied by $\eta(z)$. Taking $t=1$ in (\ref{A(t) trees}), the linear term of $\varphi_{0}(z)=(n+z)/L$ does not contribute to $W_{\varphi_{0}}^{\eta,\xi}$, and we obtain from (\ref{Q(1)})
\begin{equation}
\label{gamma(B) trees}
\gamma=-\frac{2}{L}\sum_{k=1}^{\infty}\left(\frac{B}{2}\right)^{k}\sum_{g\in\mathcal{G}_{k}}\frac{W_{1}^{\eta,\xi_{x}}(g)}{S_{t}(g)}\;.
\end{equation}
Taking the derivative at $t=1$ of equation (\ref{A(t) trees}), we obtain from (\ref{E[Q]})
\begin{equation}
\label{E(gamma) trees}
\frac{E(\gamma)-J\gamma}{p}=\frac{2(1-x)}{L(L-1)}\sum_{k=2}^{\infty}\left(\frac{B}{2}\right)^{k}\sum_{g\in\mathcal{G}_{k}}\frac{W_{z^{2}}^{\eta,\xi_{x}}(g)}{S_{t}(g)}\;,
\end{equation}
where the mean value of the current $J$ is given by $J/p=(1-x)n(L-n)/(L-1)$. We used again the fact that the linear term of $\varphi_{1}$ cancels in $W_{\varphi_{1}}^{\eta,\xi}$, while the constant term gives (\ref{gamma(B) trees}).\\\indent
Equations (\ref{gamma(B) trees}) and (\ref{E(gamma) trees}) give a parametric expression for $E(\gamma)$ similar to the one obtained for TASEP in \cite{DL98.1}. In the TASEP limit $x=0$, $\xi_{x}(z)=1$ for all $z$. Thus, from (\ref{W(g) tree 2}), neither $W_{1}^{\eta,1}(g)$ nor $W_{z^{2}}^{\eta,1}(g)$ depend on $g\in\mathcal{G}_{k}$ anymore. They are equal to
\begin{align}
W_{1}^{\eta,1}(g)_{\left|\substack{x=0}\right.}&=k\C{kL}{kn}\left/\C{L}{n}^{k}\right.\\
W_{z^{2}}^{\eta,1}(g)_{\left|\substack{x=0}\right.}&=\frac{k(k-1)n(L-n)}{kL-1}\C{kL}{kn}\left/\C{L}{n}^{k}\right.\;.\nonumber
\end{align}
From the generating function of the trees $Z_{k}(z)$ (\ref{GF trees}) at $z=1$, we recover the known parametric expression \cite{DL98.1} for $E(\gamma)$ in the TASEP limit. We also recover the large $L$ limit with non vanishing asymmetry considered in \cite{LK99.1}.
\end{section}

\begin{section}{Explicit expression for the cumulants of the current}
The parametric expression (\ref{gamma(B) trees}-\ref{E(gamma) trees}) for the generating function $E(\gamma)$ does not give directly access to the cumulants of the current. To obtain the $k$-th cumulant, one has to eliminate the parameter $B$ between equations (\ref{gamma(B) trees}) and (\ref{E(gamma) trees}). This can in fact be done systematically at all orders in $\gamma$. For this purpose, we have to introduce other combinatorial objects: forests, that is sets of trees.\\\indent
We call $\mathcal{H}$ the set of forests with trees elements of $\mathcal{G}$ of size strictly larger than $1$. We call $\widetilde{\mathcal{H}}$ the set of forests with trees elements of $\widetilde{\mathcal{G}}$ of size strictly larger than $1$ and with edges relabeled by $i_{1}$, $i_{2}$, $\ldots$ such that all the labels are different. The size $|h|$ of a forest $h$ is defined to be the sum of the sizes of the trees it contains. The number of trees in a forest $h$ will be denoted by $\overline{h}$. We call $\mathcal{H}_{r}$ (respectively $\widetilde{\mathcal{H}}_{r}$) the subset of $\mathcal{H}$ (resp. $\widetilde{\mathcal{H}}$) with forests $h$ such that $|h|-\overline{h}=r$. For $h\in\widetilde{\mathcal{H}}_{r}$, the number of edges in $h$ is equal to $r$.\\\indent
The various functions defined on $\mathcal{G}$ and $\widetilde{\mathcal{G}}$ (\text{e.g.} the sets of composite nodes $c$, elementary nodes $e$, outer edges $o$, the application $\theta$, \ldots) extend naturally to $\mathcal{H}$ and $\widetilde{\mathcal{H}}$. The operator $*$ is defined for a forest $h\in\widetilde{\mathcal{H}}$ as $h^{*}=\{g^{*},g\in h\}$. The functions $U_{\varphi,\eta}$, $V_{\xi}$ and $W_{\varphi}^{\eta,\xi}$ are simply extended to forests $h$ by replacing the $e(g)$ and $o(g)$ by the corresponding $e(h)$ and $o(h)$. Both expressions (\ref{W(g) tree 1}) and (\ref{W(g) tree 2}) for $W_{\varphi}^{\eta,\xi}$ are still equivalent in the case of forests. We will also need a symmetry factor for forests. For $h\in\mathcal{H}$, we define
\begin{equation}
S_{f}(h)=P_{f}(h)\frac{(-1)^{\overline{h}}}{(|h|-1)!}\prod_{g\in h}\frac{S_{t}(g)}{|g|}\;,
\end{equation}
where $P_{f}(h)$ is the number of permutations of the identical trees in the forest $h$.\\\indent
The equation (\ref{gamma(B) trees}) for $\gamma$ in terms of $B$ can be inverted by using the Lagrange inversion formula (see \textit{e.g.} \cite{FS09.1}). The sum over trees in equation (\ref{E(gamma) trees}) becomes then a sum over forests, and we finally obtain
\begin{equation}
\label{E(gamma) forests}
\frac{E(\gamma)-J\gamma}{p}=-\frac{2(1-x)}{L(L-1)}\sum_{r=2}^{\infty}\frac{(-L\gamma)^{r}}{2^{r}r!}\sum_{h\in\mathcal{H}_{r-1}}\frac{W_{z^{2}}^{\eta,\xi_{x}}(h)}{S_{f}(h)}\;.
\end{equation}
This equation for $E(\gamma)$ was checked by solving the functional equation (\ref{ebQR}) perturbatively (by the method described in \cite{P08.1}) up to order $7$ in $\gamma$ for all the systems with $2\leq L\leq 12$ and $n\leq L/2$. The cases $n>L/2$ then follow from the particle-hole symmetry of the system, which is verified by equation (\ref{E(gamma) forests}). Equation (\ref{E(gamma) forests}) immediately gives the known results \cite{P08.1} for the three first cumulants, and also expressions for the higher cumulants. Up to order $4$ in $\gamma$, we have
\begin{align}
\frac{(L-1)E(\gamma)}{p(1-x)}=&n(L-n)\gamma+\frac{L\gamma^{2}}{4}W\!\left[\GGa\right]+\frac{L^{2}\gamma^{3}}{72}\left(W\!\left[\GGGb\right]-9W\!\left[\GGGa\right]+9W\!\left[\GGa,\GGa\right]\right)\nonumber\\
&+\frac{L^{3}\gamma^{4}}{48}\left(W\!\left[\GGGGb\right]+3W\!\left[\GGGGa\right]-W\!\left[\GGGGc\right]\right.\nonumber\\
&\qquad\qquad\left.\textcolor{white}{\GGGGb}+W\!\left[\GGa,\GGGb\right]-9W\!\left[\GGa,\GGGa\right]+5W\!\left[\GGa,\GGa,\GGa\right]\right)\nonumber\\
&+\O{\gamma^{5}}\;,
\end{align}
with $W=W_{z^{2}}^{\eta,\xi_{x}}$.\\\indent
In the large system size limit with finite density $\rho=n/L$, the expression (\ref{E(gamma) forests}) for $E(\gamma)$ simplifies. In the scaling $1-x\sim2\Phi/\sqrt{L\rho(1-\rho)}$, Stirling's formula gives the leading order in the size $L$ of the system for all the cumulants of the current:
\begin{equation}
\frac{E(\gamma)-J\gamma}{p}=-\frac{4\Phi}{L^{2}}\sum_{r=2}^{\infty}\frac{\left(-\gamma\sqrt{\rho(1-\rho)L^{3}}\right)^{r}}{2^{r}r!}\sum_{h\in\mathcal{H}_{r-1}}\frac{\hat{W}_{z^{2}}^{e^{-z^{2}/2},\tanh(|z|\Phi)}(h)}{S_{f}(h)}\;.
\end{equation}
Here, $\hat{W}$ is defined in the same way as $W$ except for the fact that the discrete sums over the indices $i_{j}$ are replaced by integrals between $-\infty$ and $+\infty$.
\end{section}

\begin{section}{Conclusion}
With our conjecture (\ref{gamma(B) trees}-\ref{E(gamma) trees}) for the generating function of the cumulants of the current, we recover the exact result by Derrida and Lebowitz for the totally asymmetric model \cite{DL98.1}, as well as the expression obtained by Lee and Kim for the partially asymmetric model with non vanishing asymmetry \cite{LK99.1}. The expression (\ref{E(gamma) forests}) for $E(\gamma)$ then gives explicit expressions for the cumulants of the current, and we recover the known results for the three first cumulants \cite{P08.1}. It would be interesting to calculate from (\ref{E(gamma) forests}) the cumulants of the current for the weakly asymmetric model, where the asymmetry $1-x$ scales as $1/L$. It should be possible to recover the expression obtained in \cite{PM09.1} in terms of Bernoulli numbers. So far, we only checked this for the four first cumulants.\\\indent
A distinctive feature between the known result for TASEP \cite{DL98.1} and the expression with partial asymmetry (\ref{E(gamma) forests}) is the appearance of tree structures. This emphasizes the importance of combinatorics in theoretical physics, in particular in relation with integrable models. It would be interesting to know whether tree structures also appear in other situations, in particular for the open ASEP, the multispecies ASEP and the ASEP on the infinite line.

\subsection*{Acknowledgments}
It is a pleasure to thank Olivier Golinelli and Kirone Mallick for many useful discussions.
\end{section}


\end{document}